\documentstyle[11pt,epsf]{article}
\makeatletter\@addtoreset {equation}{section}\makeatother

\begin{document}
\title{\bf Eigenfunctions of linearized integrable equations expanded 
around an arbitrary solution}
\author{Jianke Yang \\ 
Department of Mathematics and Statistics \\ 
University of Vermont \\
Burlington, VT 05401, USA \\
Tel: 802-6564314;  \hspace{0.1cm} Fax: 802-6562552 \\
Email: jyang@emba.uvm.edu }

\date{}
\maketitle

{\bf Abstract}

Complete eigenfunctions of linearized integrable equations expanded around an 
arbitrary solution are obtained for the 
Ablowitz-Kaup-Newell-Segur (AKNS) hierarchy and 
the Korteweg-de Vries (KdV) hierarchy. 
It is shown that the linearization operators and the recursion 
operator which generates the hierarchy are commutable. Consequently, 
eigenfunctions of the linearization operators are precisely squared
eigenfunctions of the associated eigenvalue problem. 
Similar results are obtained for the adjoint linearization operators as well. 
These results make a simple connection between the direct 
soliton/multi-soliton perturbation
theory and the inverse-scattering based perturbation theory for these hierarchy 
equations. 

\section{Introduction}
Integrable equations are nonlinear evolution equations which can be solved
exactly by the inverse scattering method. Over the past
few decades, it has been discovered that many physically important equations
such as the Korteweg-de Vries (KdV), nonlinear Schr\"odinger (NLS) and
sine-Gordon equations are integrable (see \cite{ablowitz_segur} and the 
references therein). Linearization of an integrable equation around its
solution arises in many important applications, most notably in 
a direct soliton/multi-soliton perturbation theory. 
In such situations, eigenfunctions of linearization operators and their
completeness are the key questions. 
For linearization around single-soliton
solutions, these complete eigenfunctions have been obtained for a large class
of integrable equations such as the KdV hierarchy, NLS hierarchy, 
modified-KdV hierarchy, sine-Gordon, and Benjamin-Ono equations
\cite{keener1977, fogel1977, 
chen1980, kaup1990, herman1990, yan1996, matsuno1997, yang2000}. 
It has been found that these eigenfunctions are related to squared
eigenfunctions of the associated eigenvalue problem (except for the Benjamin-Ono equation). 
However, for linearization around a general solution such as a multi-soliton
solution, complete eigenfunctions are known for much less integrable equations 
\cite{keener1977, matsuno1997}. Some general ideas have been proposed 
to determine these eigenfunctions though. 
One idea by Keener and McLaughlin \cite{keener1977} is that eigenfunctions 
of a linearization operator expanded around an arbitrary solution are the 
variations of the solution with respect to each parameter in the scattering data. 
Another idea by Herman \cite{herman1990} is to utilize the Lax pair
of the integrable equation and find special combinations of squared eigenfunctions
of the associated eigenvalue problem, so that these combinations satisfy
the linearized equation of the evolution equation. But in both approaches, 
each equation has to be treated separately. In addition, for each equation,
much work is needed to find eigenfunctions of the linearization operator, 
or relate them to squared eigenfunctions of the associated eigenvalue problem. 
The idea by Yang \cite{yang2000}, however, is free of these problems. 
This idea  is to show that linearization operators
of a hierarchy and the recursion operator which generates the hierarchy
are commutable, thus they share the same set of eigenfunctions. Furthermore, these
eigenfunctions are simply squared eigenfunctions of the associate eigenvalue
problem. Compared to the other two approaches, this method explicitly gives the eigenfunctions
of linearization operators in the simplest way.  In addition, it treats an entire
hierarchy all at once. In \cite{yang2000}, this idea was applied only 
to linearizations of the KdV, NLS and modified-KdV hierarchies 
around single-soliton solutions. In that special case, 
our results went beyond commutability of the operators. We also showed that
linearization operators of the hierarchy equations could be factored into
the recursion operator of the hierarchy and 
the linearization operator of the lowest-order equation in the hierarchy. 
Commutability between linearization operators and the recursion operator
is a simple consequence of this factorization representation for the 
linearization operators. 

In this article, we extend the results of Yang \cite{yang2000} to the 
Ablowitz-Kaup-Newell-Segur (AKNS) hierarchy and KdV hierarchy linearized
around an {\em arbitrary} solution. The AKNS hierarchy is the family of integrable
equations associated with the Zakharov-Shabat eigenvalue problem
\cite{ZS, AKNS}, and the KdV hierarchy is associated with 
the Schr\"odinger eigenvalue problem \cite{AKNS,GGKM}.
In this general case, we can still show that 
linearization operators are commutable with the recursion operator 
of the hierarchy (the factorization result of linearization operators
for single-soliton solutions does not extend to this general case
though). This commutability
allows us to establish that complete eigenfunctions of linearization operators
in the AKNS or KdV hierarchy are simply squared eigenfunctions of the Zakharov-Shabat
or Schr\"odinger operator. Similar results can be obtained for adjoint linearization
operators as well. 
In a direct soliton/multi-soliton perturbation theory, these squared eigenfunctions
will then serve as the expansion basis for perturbation solutions. 
Interestingly, these same squared eigenfunctions were also used to 
expand perturbation solutions in the inverse-scattering based perturbation theory
\cite{kaup1976, kaupnewell1978, newell1980}. 
Thus, our results in this article indicate that, at a deeper level, 
the direct soliton/multi-soliton perturbation theory and the 
inverse-scattering based perturbation theory are really equivalent. 

\section{Eigenfunctions of linearization operators for the AKNS hierarchy}

The AKNS hierarchy associated with the Zakharov-Shabat eigenvalue problem is
\cite{AKNS}: 
\begin{equation}\label{rq}
\left[ \begin{array}{c} r_t \\ -q_t \end{array} \right]
+i\omega(2L_z^+) \left[ \begin{array}{c} r \\ q \end{array} \right]=0, 
\end{equation}
where the recursion operator $L_z^+$ is
\begin{equation} \label{Lz+}
L_z^+=\frac{1}{2i}\left[ \begin{array}{cc} 
\frac{\partial}{\partial x}-2r\int_{-\infty}^x dy q   &  2r\int_{-\infty}^xdy r \\
-2q\int_{-\infty}^xdy q   & -\frac{\partial}{\partial x}+2q\int_{-\infty}^x dy r
 \end{array} \right],
\end{equation}
and $\omega(k)$ is the dispersion relation of the linear equation in the 
$r$-component. In this section, we require that 
$\omega(k)$ is an entire function. 
When $\omega(k)=k^2$ and $q=-r^*$, 
Eq. (\ref{rq}) reduces to the NLS equation; when $\omega(k)=k^3$ and $q=-r$, 
the modified-KdV equation results.
The adjoint operator of $L_z^+$ is
\begin{equation} \label{Lz}
L_z=\frac{1}{2i}\left[ \begin{array}{cc}
-\frac{\partial}{\partial x}-2q\int^{\infty}_x dy r   &  -2q\int^{\infty}_xdy q \\
2r\int^{\infty}_xdy r  & \frac{\partial}{\partial x}+2r\int^{\infty}_x dy q
 \end{array} \right].
\end{equation}

Now suppose $[r_0(x, t), q_0(x, t)]^T$ is an arbitrary solution of the
evolution equation (\ref{rq}). To avoid dealing with divergent integrals
in the following analysis, we require that this solution vanish as
$|x|$ goes to infinity. But generalization to non-vanishing solutions
is also possible by appropriately defining divergent integrals, as we did
in \cite{yang2000}. Next we linearize Eq. (\ref{rq}) around this arbitrary
solution. For this purpose, we write 
\begin{equation} \label{rqlinearize}
r=r_0(x, t)+\tilde{r}(x, t), \hspace{0.5cm}
q=q_0(x, t)-\tilde{q}(x, t), 
\end{equation}
where $\tilde{r}, \tilde{q} \ll 1$. Note that we deliberately
introduced opposite signs in front of $r$ and $q$'s perturbations. 
This is important for obtaining the commutability relations which we will
present later in this section. 
When Eq. (\ref{rqlinearize}) is substituted into Eq. (\ref{rq}), 
linearization of Eq. (\ref{rq}) is: 
\begin{equation}
L \left(\begin{array}{c} \tilde{r} \\ \tilde{q} \end{array} \right)=0, 
\end{equation}
where $L$ is the linearization operator. We denote the adjoint operator
of $L$ as $L^A$. We also denote $L_0^+$ and $L_0$ as the 
recursion operators $L_z^+$ and $L_z$ with $r$ and $q$ 
replaced by the solutions $r_0(x, t)$ and $q_0(x, t)$. 
The primary objective of this section is to show that operators
$L$ and $L_0^+$ commute, and $L^A$ and $L_0$ commute, i.e., 
\begin{equation} \label{comm1}
LL_0^+=L_0^+L, 
\end{equation}
and 
\begin{equation} \label{comm2}
L^AL_0=L_0 L^A. 
\end{equation}
We will prove relation (\ref{comm1}) first. Relation (\ref{comm2}) will then follow
naturally. 

Without loss of generality, we assume that the dispersion relation
$\omega(k)$ is a power function, $\omega(k)=k^n$, where $n$ is a non-negative
integer. The reason is that any entire function of $\omega(k)$ can 
be expanded into a power series. Linearization of
operator $L_z^+$ around the solution $(r_0, q_0)$ is
\begin{equation} \label{Lz+linearize}
L_z^+=L_0^+ +\frac{1}{2i}{\mathcal{F}}
\left[ \begin{array}{c} \tilde{r} \\ \tilde{q} \end{array} \right]
+O(\tilde{r}^2, \tilde{r}\tilde{q}, \tilde{q}^2), 
\end{equation}
where the operator $\mathcal{F}$ is defined as 
\begin{equation} 
\mathcal{F}\left[ \begin{array}{c} \tilde{r} \\ \tilde{q} \end{array} \right]
=\left( \begin{array}{cc}
-2\tilde{r}\int_{-\infty}^x dy q_0+2r_0\int_{-\infty}^x dy \tilde{q} &
2\tilde{r}\int_{-\infty}^x dy r_0+2 r_0\int_{-\infty}^x dy \tilde{r} \\
2\tilde{q}\int_{-\infty}^x dy q_0+2 q_0\int_{-\infty}^x dy \tilde{q} & 
-2\tilde{q}\int_{-\infty}^x dy r_0+2q_0\int_{-\infty}^x dy \tilde{r}
\end{array}\right).
\end{equation}
Then, for power functions of $\omega(k)$, the linearization operator $L$ is 
simply: 
\begin{equation} \label{L}
L\left( \begin{array}{c} \tilde{r} \\ \tilde{q} \end{array} \right)
=\left( \begin{array}{c} \tilde{r} \\ \tilde{q} \end{array} \right)_t
+i(2L_0^+)^n \left( \begin{array}{c} \tilde{r} \\ -\tilde{q} \end{array} \right)
+\sum_{k=1}^n (2L_0^+)^{k-1}
{\mathcal{F}}\left[ \begin{array}{c} \tilde{r} \\ \tilde{q} \end{array} \right]
(2L_0^+)^{n-k}
\left( \begin{array}{c} r_0 \\ q_0 \end{array} \right). 
\end{equation}
Denoting 
\begin{equation}
\left( \begin{array}{c} P_n \\ Q_n \end{array} \right)
=-i (2L_0^+)^n \left( \begin{array}{c} r_0 \\ q_0 \end{array} \right),
\end{equation}
then the evolution of $(r_0, q_0)$ becomes 
\begin{equation}\label{PQ}
\left( \begin{array}{c} r_{0} \\ -q_{0} \end{array} \right)_t
=\left( \begin{array}{c} P_n \\ Q_n \end{array} \right), 
\end{equation}
and the functions $(P_n, Q_n)$ satisfy the recursion relation
\begin{equation} \label{recursion}
\left( \begin{array}{c} P_{n+1} \\ Q_{n+1} \end{array} \right)
=2L_0^+ \left( \begin{array}{c} P_n \\ Q_n \end{array} \right).
\end{equation}
To establish the commutability of operators $L$ and $L_0^+$, we examine
the function
\begin{equation}
{\mathcal{H}}_n \equiv 2i (L_0^+ L-L L_0^+)
\left( \begin{array}{c} \tilde{r} \\ \tilde{q} \end{array} \right). 
\end{equation}
Simple calculations show that ${\mathcal{H}}_n$ has the following expression: 
\begin{eqnarray} \label{Hn}
{\mathcal{H}}_n= & 
\left( \begin{array}{c} 2P_n\int_{-\infty}^x(q_0 \tilde{r}-r_0\tilde{q})dy
			-2r_0 \int_{-\infty}^x(Q_n \tilde{r}+P_n\tilde{q})dy \\
			-2Q_n\int_{-\infty}^x(q_0 \tilde{r}-r_0\tilde{q})dy
			-2q_0 \int_{-\infty}^x(Q_n \tilde{r}+P_n\tilde{q})dy 
	\end{array}\right) \nonumber \\
& -(2L_0^+)^{n+1}\left( \begin{array}{c} \tilde{r} \\ -\tilde{q} \end{array} \right)
+(2L_0^+)^{n}\left( \begin{array}{cc} 1 & 0 \\ 0 & -1 \end{array} \right)
(2L_0^+)\left( \begin{array}{c} \tilde{r} \\ \tilde{q} \end{array} \right) \nonumber \\
& +\sum_{k=1}^n (2L_0^+)^{k-1} \left\{2iL_0^+ {\mathcal{F}} 
\left[ \begin{array}{c} \tilde{r} \\ \tilde{q} \end{array} \right]
-{\mathcal{F}}
\left[2iL_0^+\left( \begin{array}{c} \tilde{r} \\ \tilde{q} \end{array} \right)\right]\right\}
(2L_0^+)^{n-k}\left( \begin{array}{c} r_0 \\ q_0 \end{array} \right). 
\end{eqnarray}
Here we have replaced the time derivatives $r_{0t}$ and $-q_{0t}$ in 
${\mathcal{H}}_n$ by $P_n$ and $Q_n$ in view of Eq. (\ref{PQ}). 
It is important to realize that the above ${\mathcal{H}}_n$ expression (\ref{Hn})
is now purely algebraic, and is independent of the evolution equation (\ref{PQ}). 
Below we will use algebraic manipulations and the induction method to prove that
${\mathcal{H}}_n$ is zero for all $n\ge 0$. 

When $n=0$ or 1, one can verify directly that ${\mathcal{H}}_n$ is indeed zero. 
Now we assume that ${\mathcal{H}}_n=0$ for some $n\ge 0$. Then we try to show that
${\mathcal{H}}_{n+1}=0$. For this purpose, we calculate the quantity
${\mathcal{H}}_{n+1}-2L_0^+ {\mathcal{H}}_n$. It turns out that most of the summation
terms in ${\mathcal{H}}_{n+1}$ and $2L_0^+ {\mathcal{H}}_n$ cancel each other out. 
The terms left over are 
\begin{eqnarray} \label{Hn+1}
{\mathcal{H}}_{n+1}-2L_0^+ {\mathcal{H}}_n= & 
\left( \begin{array}{c} 2P_{n+1}\int_{-\infty}^x(q_0 \tilde{r}-r_0\tilde{q})dy
                        -2r_0 \int_{-\infty}^x(Q_{n+1} \tilde{r}+P_{n+1}\tilde{q})dy \\
                        -2Q_{n+1}\int_{-\infty}^x(q_0 \tilde{r}-r_0\tilde{q})dy
                        -2q_0 \int_{-\infty}^x(Q_{n+1} \tilde{r}+P_{n+1}\tilde{q})dy
        \end{array}\right) \nonumber \\
& -2L_0^+ \left( \begin{array}{c} 2P_n\int_{-\infty}^x(q_0 \tilde{r}-r_0\tilde{q})dy
                        -2r_0 \int_{-\infty}^x(Q_n \tilde{r}+P_n\tilde{q})dy \\
                        -2Q_n\int_{-\infty}^x(q_0 \tilde{r}-r_0\tilde{q})dy
                        -2q_0 \int_{-\infty}^x(Q_n \tilde{r}+P_n\tilde{q})dy
        \end{array}\right) \nonumber \\
& -\left\{2L_0^+{\mathcal{F}}\left[ \begin{array}{c} \tilde{r} \\ \tilde{q} \end{array} \right]
+i{\mathcal{F}}\left[2iL_0^+\left( \begin{array}{c} \tilde{r} \\ \tilde{q} \end{array} \right)\right]
\right\}
\left( \begin{array}{c} P_n \\ Q_n \end{array} \right). 
\end{eqnarray}
When the recursion relation (\ref{recursion}) for $P_{n+1}$ and $Q_{n+1}$ 
is substituted into the above expression, algebraic simplifications immediately
reveal that 
\begin{equation}
{\mathcal{H}}_{n+1}-2L_0^+ {\mathcal{H}}_n= 0. 
\end{equation}
Since ${\mathcal{H}}_n$ is zero by assumption, it then follows that ${\mathcal{H}}_{n+1}=0$. 
Thus ${\mathcal{H}}_n=0$ for all $n\ge 0$, which means that 
$L$ and $L_0^+$ are commutable. For a general entire function
of the dispersion relation $\omega(k)$, this result still holds, as
an entire function can be expanded into a power series. 

The proof for the commutability of $L^A$ and $L_0$ is trivial once 
the commutability of $L$ and $L_0^+$ has been established. 
The adjoint operator of $LL_0^+$ is $L_0L^A$, and the adjoint
of $L_0^+L$ is $L^AL_0$. Since $LL_0^+=L_0^+L$, their adjoints
are certainly the same, i.e., $L_0L^A=L^AL_0$. Thus 
$L^A$ and $L_0$ are also commutable. 

An important consequence of the commutability relations (\ref{comm1}) and
(\ref{comm2}) is that $L$ ($L^A$) and $L_0^+$ ($L_0$) share the same set of
eigenfunctions. To see how this comes about, let us assume that 
$\Psi(x, t, \zeta)$ is a continuous eigenfunction of $L_0^+$ with real eigenvalue
$\zeta$, i.e., 
\begin{equation}
L_0^+\Psi=\zeta \Psi. 
\end{equation}
Under the condition that $[r_0(x, t), q_0(x, t)]$ vanishes as $|x|$ goes to infinity, 
we can impose the boundary condition for $\Psi$ as 
\begin{equation} \label{Psibc}
\Psi(x, t, \zeta) \longrightarrow
\left( \begin{array}{c} 0 \\ -e^{-2i\zeta x} \end{array}\right), 
\hspace{0.5cm} x \rightarrow -\infty. 
\end{equation}
Since $L$ and $L_0^+$ are commutable, we have
\begin{equation}
L_0^+ L\Psi=\zeta L\Psi. 
\end{equation}
Thus $L\Psi$ is also an eigenfunction of $L_0^+$ with eigenvalue $\zeta$. 
As $x$ goes to infinity, the linearization operator $L$ becomes
\begin{equation} \label{Lbc}
L \longrightarrow \left( \begin{array}{cc}
\partial_t+i\omega(-i\partial_x) & 0 \\
0 & \partial_t-i\omega(i\partial_x) \end{array} \right), 
\hspace{0.5cm} |x| \rightarrow \infty. 
\end{equation}
Consequently,  the boundary condition for $L\Psi$ can be obtained from Eqs. (\ref{Psibc}) and
(\ref{Lbc}) as 
\begin{equation}
L\Psi(x, t, \zeta)\longrightarrow 
 -i\omega(2\zeta) \left( \begin{array}{c} 0 \\ -e^{-2i\zeta x} \end{array}\right),
\hspace{0.5cm} x \rightarrow -\infty, 
\end{equation}
which is proportional to the boundary condition (\ref{Psibc}) of eigenfunction $\Psi$. 
Then it becomes clear that $L\Psi$ and $\Psi$ are the same eigenfunction of
operator $L_0^+$ with eigenvalue $\zeta$ (i.e., they are linearly dependent). 
In view of their boundary conditions, we see that
\begin{equation}
L\Psi=-i\omega(2\zeta)\Psi, 
\end{equation}
i.e., $\Psi(x, t, \zeta)$ is also a continuous eigenfunction of operator $L$ with
eigenvalue $-i\omega(2\zeta)$. 

For the same real eigenvalue $\zeta$, $L_0^+$ has another linearly independent
eigenfunction $\bar{\Psi}$ with boundary condition
\begin{equation}
\bar{\Psi}(x, t, \zeta) \longrightarrow
\left( \begin{array}{c} e^{2i\zeta x} \\ 0 \end{array}\right),
\hspace{0.5cm} x \rightarrow -\infty.
\end{equation}
Similar analysis shows that $\bar{\Psi}$ is also a continuous eigenfunction of $L$, but
with eigenvalue $i\omega(2\zeta)$, i.e., 
\begin{equation}
L\bar{\Psi}=i\omega(2\zeta)\bar{\Psi}. 
\end{equation}
For the discrete eigenfunctions and generalized eigenfunctions of $L_0^+$, 
same analysis indicates that they are also discrete eigenfunctions and generalized eigenfunctions
of $L$. Thus $L_0^+$ and $L$ indeed share the same set of eigenfunctions. Naturally, 
the same statement applies to $L_0$ and $L_A$ as well.

What exactly are the sets of eigenfunctions for $L$ and $L^A$? Are these sets
complete? In view of our results above, we only need to find the answers
for operators $L_0^+$ and $L_0$. The eigenfunctions for $L_0^+$ and $L_0$
and their closure have been known for over twenty years from the celebrated work 
by Ablowitz, Kaup, Newell and Segur \cite{AKNS} and by 
Kaup \cite{kaup1976b}. The results can be summarized as follows. 

Consider the Zakharov-Shabat eigenvalue problem with potential 
$[q_0(x, t), r_0(x, t)]$: 
\begin{equation} \label{v1}
v_{1x}+i\zeta v_1=q_0(x, t) v_2, 
\end{equation}
\begin{equation} \label{v2}
v_{2x}-i\zeta v_2=r_0(x, t) v_1, 
\end{equation}
and define Jost functions for real $\zeta$ as 
\begin{equation}
\psi(x, t, \zeta) = \left[\begin{array}{c}\psi_1 \\ \psi_2 \end{array}\right]
\longrightarrow \left[\begin{array}{c} 0 \\ 1 \end{array}\right]e^{i\zeta x}, 
\hspace{0.7cm} x \rightarrow \infty, 
\end{equation}
\begin{equation}
\bar{\psi}(x, t, \zeta) = \left[\begin{array}{c}\bar{\psi}_1 \\ \bar{\psi}_2 
\end{array}\right]
\longrightarrow \left[\begin{array}{c} 1 \\ 0 \end{array}\right]e^{-i\zeta x}, 
\hspace{0.7cm} x \rightarrow \infty, 
\end{equation}
\begin{equation}
\phi(x, t, \zeta) = \left[\begin{array}{c}\phi_1 \\ \phi_2 \end{array}\right]
\longrightarrow \left[\begin{array}{c} 1 \\ 0 \end{array}\right]e^{-i\zeta x},
\hspace{0.7cm} x \rightarrow -\infty,
\end{equation}
\begin{equation}
\bar{\phi}(x, t, \zeta) = \left[\begin{array}{c}\bar{\phi}_1 \\ \bar{\phi}_2
\end{array}\right]
\longrightarrow \left[\begin{array}{c} 0 \\ -1 \end{array}\right]e^{i\zeta x}, 
\hspace{0.7cm} x \rightarrow -\infty. 
\end{equation}
The right and left solutions are related by 
\begin{equation} \label{phiphibar1}
\phi(x, t, \zeta)=a(t, \zeta)\bar{\psi}(x, t, \zeta)+b(t, \zeta)\psi(x, t, \zeta), 
\end{equation}
\begin{equation}\label{phiphibar2}
\bar{\phi}(x, t, \zeta)=-\bar{a}(t, \zeta)\psi(x, t, \zeta)+
\bar{b}(t, \zeta)\bar{\psi}(x, t, \zeta), 
\end{equation}
where 
\begin{equation} \label{wronskian}
\bar{a}(t, \zeta) a(t, \zeta)+\bar{b}(t, \zeta)b(t, \zeta)=1
\end{equation}
from Wronskian relations. With Eq. (\ref{wronskian}), the inverse of 
Eqs. (\ref{phiphibar1}) and (\ref{phiphibar2}) is 
\begin{equation}
\psi=-a\bar{\phi}+\bar{b}\phi, 
\end{equation}
\begin{equation}
\bar{\psi}=\bar{a}\phi+b\bar{\phi}, 
\end{equation}
where we have suppressed the dependent variables $x, t$ and $\zeta$. In addition to 
the continuous spectrum ($\zeta$ real), Eqs. (\ref{v1}) and (\ref{v2}) may also 
possess discrete eigenvalues (bound states) in the upper and the lower 
half $\zeta$-plane. In the upper half plane, these occur whenever 
$a(t, \zeta)=0$, and we designate them by $\zeta_k$, $k=1, 2, \dots, N$, where
$N$ is the total number of bound states in the upper half $\zeta$-plane. 
At $\zeta=\zeta_k$, $\phi$ and $\psi$ become linearly dependent and
\begin{equation}
\phi(x, t, \zeta_k)=b(t, \zeta_k)\psi(x, t, \zeta_k), \hspace{0.5cm}
k=1, 2, \dots, N. 
\end{equation}
In the lower half $\zeta$-plane, bound states correspond to zeros of $\bar{a}(t, \zeta)$
which we designate by $\bar{\zeta}_k, k=1, 2, \dots, \bar{N}$. At 
$\zeta=\bar{\zeta}_k$, 
\begin{equation}
\bar{\phi}(x, t, \bar{\zeta}_k)=\bar{b}(t, \bar{\zeta}_k)\bar{\psi}(x, t, \bar{\zeta}_k), 
\hspace{0.5cm} k=1, 2, \dots, \bar{N}. 
\end{equation}
It is important to note that when $[q_0(x, t), r_0(x, t)]$ is a solution of the
AKNS hierarchy (\ref{rq}), the discrete eigenvalues $\zeta_k$ and $\bar{\zeta}_k$
are independent of time $t$. 

With the above notations, 
the eigenfunctions and generalized eigenfunctions of operators $L_0^+$ and $L_0$ are 
simply squared eigenstates of the Zakharov-Shabat system (\ref{v1}) and (\ref{v2}) 
\cite{AKNS}. 
Specifically, the set of eigenfunctions and generalized eigenfunctions 
for $L_0^+$ is
\small
\begin{equation}\label{aknsset1}
\left\{ \left[\begin{array}{c}\phi_2^2 \\ -\phi_1^2 \end{array}\right]_\zeta, 
\left[\begin{array}{c}\bar{\phi}_2^2 \\ -\bar{\phi}_1^2\end{array}\right]_\zeta, 
\zeta \; \mbox{real}; 
\left[\begin{array}{c}\phi_2^2 \\ -\phi_1^2 \end{array}\right]_{\zeta_k}, 
\frac{\partial}{\partial\zeta}
\left[\begin{array}{c}\phi_2^2 \\ -\phi_1^2 \end{array}\right]_{\zeta_k}, 
1\le k \le N; 
\left[\begin{array}{c}\bar{\phi}_2^2 \\ -\bar{\phi}_1^2\end{array}\right]_{\bar{\zeta}_k},
 \frac{\partial}{\partial\zeta}
\left[\begin{array}{c}\bar{\phi}_2^2 \\ -\bar{\phi}_1^2\end{array}\right]_{\bar{\zeta}_k}, 
1\le k \le \bar{N}
\right\}, 
\end{equation}
\normalsize
and the set of such eigenfunctions for $L_0$ is
\small
\begin{equation} \label{aknsset2}
\left\{\left[\begin{array}{c}\psi_1^2 \\ \psi_2^2 \end{array}\right]_\zeta, 
\left[\begin{array}{c}\bar{\psi}_1^2 \\ \bar{\psi}_2^2\end{array}\right]_\zeta, \zeta \; \mbox{real}; 
\left[\begin{array}{c}\psi_1^2 \\ \psi_2^2 \end{array}\right]_{\zeta_k}, 
\frac{\partial}{\partial\zeta}
\left[\begin{array}{c}\psi_1^2 \\ \psi_2^2 \end{array}\right]_{\zeta_k},
1\le k \le N;
\left[\begin{array}{c}\bar{\psi}_1^2 \\ \bar{\psi}_2^2\end{array}\right]_{\bar{\zeta}_k}, 
\frac{\partial}{\partial\zeta}
\left[\begin{array}{c}\bar{\psi}_1^2 \\ \bar{\psi}_2^2\end{array}\right]_{\bar{\zeta}_k}, 
1\le k \le \bar{N}\right\}. 
\end{equation}
\normalsize
It has been shown by Kaup \cite{kaup1976b} that each of these two sets
is complete. The orthogonality and
inner products of functions in these sets have also been obtained there. 
In view of these facts, we then conclude that 
the sets (\ref{aknsset1}) and (\ref{aknsset2}) are also 
the complete sets of eigenfunctions and generalized eigenfunctions
for linearization operators $L$ and $L^A$ respectively. What about
the corresponding eigenvalues? The eigenvalues are actually quite easy 
to obtain from the asymptotic behaviors of these eigenfunctions. For operator $L$, 
the results are: 
\begin{equation}
L\left[\begin{array}{c}\phi_2^2 \\ -\phi_1^2 \end{array}\right]_\zeta=-i\omega(2\zeta)
\left[\begin{array}{c}\phi_2^2 \\ -\phi_1^2 \end{array}\right]_\zeta, \hspace{0.5cm}
\zeta \; \mbox{real}; 
\end{equation}
\begin{equation}
L\left[\begin{array}{c}\bar{\phi}_2^2 \\ -\bar{\phi}_1^2\end{array}\right]_\zeta=i\omega(2\zeta)
\left[\begin{array}{c}\bar{\phi}_2^2 \\ -\bar{\phi}_1^2\end{array}\right]_\zeta, 
\hspace{0.5cm} \zeta\; \mbox{real}; 
\end{equation}
\begin{equation}
L \left[\begin{array}{c}\phi_2^2 \\ -\phi_1^2 \end{array}\right]_{\zeta_k}
=-i\omega(2\zeta_k) \left[\begin{array}{c}\phi_2^2 \\ -\phi_1^2 \end{array}\right]_{\zeta_k}, 
\hspace{0.5cm} 1\le k \le N;
\end{equation}
\begin{equation}
L \left[\begin{array}{c}\bar{\phi}_2^2 \\ -\bar{\phi}_1^2\end{array}\right]_{\bar{\zeta}_k}
=i\omega(2\bar{\zeta}_k) 
\left[\begin{array}{c}\bar{\phi}_2^2 \\ -\bar{\phi}_1^2\end{array}\right]_{\bar{\zeta}_k}, 
\hspace{0.5cm} 1\le k \le \bar{N};
\end{equation}
\begin{equation}
L \frac{\partial}{\partial\zeta}
\left[\begin{array}{c}\phi_2^2 \\ -\phi_1^2 \end{array}\right]_{\zeta_k}
=-i\omega(2\zeta_k) \frac{\partial}{\partial\zeta}
\left[\begin{array}{c}\phi_2^2 \\ -\phi_1^2 \end{array}\right]_{\zeta_k}
-2i\omega'(2\zeta_k)\left[\begin{array}{c}\phi_2^2 \\ -\phi_1^2 \end{array}\right]_{\zeta_k}, 
\hspace{0.5cm}1\le k \le N;
\end{equation}
and 
\begin{equation}
L \frac{\partial}{\partial\zeta}
\left[\begin{array}{c}\bar{\phi}_2^2 \\ -\bar{\phi}_1^2\end{array}\right]_{\bar{\zeta}_k}
=i\omega(2\bar{\zeta}_k) \frac{\partial}{\partial\zeta}
\left[\begin{array}{c}\bar{\phi}_2^2 \\ -\bar{\phi}_1^2\end{array}\right]_{\bar{\zeta}_k}
+2i\omega'(2\bar{\zeta}_k)
\left[\begin{array}{c}\bar{\phi}_2^2 \\ -\bar{\phi}_1^2\end{array}\right]_{\bar{\zeta}_k}, 
\hspace{0.5cm} 1\le k \le \bar{N}. 
\end{equation}
The results for $L^A$ are: 
\begin{equation}
L^A\left[\begin{array}{c}\psi_1^2 \\ \psi_2^2 \end{array}\right]_\zeta=-i\omega(2\zeta)
\left[\begin{array}{c}\psi_1^2 \\ \psi_2^2 \end{array}\right]_\zeta, \hspace{0.5cm}
\zeta \; \mbox{real}; 
\end{equation}
\begin{equation}
L^A\left[\begin{array}{c}\bar{\psi}_1^2 \\ \bar{\psi}_2^2\end{array}\right]_\zeta=i\omega(2\zeta)
\left[\begin{array}{c}\bar{\psi}_1^2 \\ \bar{\psi}_2^2\end{array}\right]_\zeta, 
\hspace{0.5cm} \zeta \;  \mbox{real}; 
\end{equation}
\begin{equation}
L^A \left[\begin{array}{c}\psi_1^2 \\ \psi_2^2 \end{array}\right]_{\zeta_k}
=-i\omega(2\zeta_k) \left[\begin{array}{c}\psi_1^2 \\ \psi_2^2 \end{array}\right]_{\zeta_k}, 
\hspace{0.5cm} 1\le k \le N;
\end{equation}
\begin{equation}
L^A \left[\begin{array}{c}\bar{\psi}_1^2 \\ \bar{\psi}_2^2\end{array}\right]_{\bar{\zeta}_k}
=i\omega(2\bar{\zeta}_k) 
\left[\begin{array}{c}\bar{\psi}_1^2 \\ \bar{\psi}_2^2\end{array}\right]_{\bar{\zeta}_k}, 
\hspace{0.5cm} 1\le k \le \bar{N};
\end{equation}
\begin{equation}
L^A \frac{\partial}{\partial\zeta}
\left[\begin{array}{c}\psi_1^2 \\ \psi_2^2 \end{array}\right]_{\zeta_k}
=-i\omega(2\zeta_k) \frac{\partial}{\partial\zeta}
\left[\begin{array}{c}\psi_1^2 \\ \psi_2^2 \end{array}\right]_{\zeta_k}
-2i\omega'(2\zeta_k)\left[\begin{array}{c}\psi_1^2 \\ \psi_2^2 \end{array}\right]_{\zeta_k}, 
\hspace{0.5cm}1\le k \le N;
\end{equation}
and 
\begin{equation}
L^A \frac{\partial}{\partial\zeta}
\left[\begin{array}{c}\bar{\psi}_1^2 \\ \bar{\psi}_2^2\end{array}\right]_{\bar{\zeta}_k}
=i\omega(2\bar{\zeta}_k) \frac{\partial}{\partial\zeta}
\left[\begin{array}{c}\bar{\psi}_1^2 \\ \bar{\psi}_2^2\end{array}\right]_{\bar{\zeta}_k}
+2i\omega'(2\bar{\zeta}_k)
\left[\begin{array}{c}\bar{\psi}_1^2 \\ \bar{\psi}_2^2\end{array}\right]_{\bar{\zeta}_k}, 
\hspace{0.5cm} 1\le k \le \bar{N}. 
\end{equation}

Lastly, we note that in the development of a direct soliton/multi-soliton perturbation
theory, it is often convenient to use the derivatives of soliton/multi-soliton
solutions with respect to soliton parameters as discrete eigenfunctions and 
generalized eigenfunctions of the linearization operator \cite{keener1977, kaup1990, yang2000}. 
These derivative states span the same linear space as the discrete eigenfunctions in 
the set (\ref{aknsset1}) do. Thus use of either discrete set is sufficient. 

\section{Eigenfunctions of linearization operators for the KdV hierarchy}
For the KdV hierarchy, similar results hold. The analysis is simpler though as we only 
have a scaler equation to consider. This hierarchy can be written as \cite{AKNS}: 
\begin{equation} \label{q}
q_t+C(4L_s^+)q_x=0,   
\end{equation}
where $q(x, t)$ is a real function, $C(k^2)$ is the phase velocity of the linear equation, 
and the recursion operator $L_s^+$ is: 
\begin{equation} \label{Ls+}
L_s^+=-\frac{1}{4}\frac{\partial^2}{\partial x^2}-q+\frac{1}{2}q_x \int_x^\infty dy. 
\end{equation}
Here the subscript ``$s$'' in $L_s^+$ refers to ``Schr\"odinger'', as
the associated eigenvalue problem for the KdV hierarchy (\ref{q})
is the Schr\"odinger equation \cite{AKNS,GGKM}.
In this section, we require the phase velocity function $C(z)$ to be entire. 
The adjoint operator of $L_s^+$ is: 
\begin{equation} \label{Ls}
L_s=-\frac{1}{4}\frac{\partial^2}{\partial x^2}-q+\frac{1}{2} \int_{-\infty}^x dy q_y. 
\end{equation}

Suppose $q_0(x, t)$ is an arbitrary solution of the evolution equation (\ref{q}). 
To linearize Eq. (\ref{q}) around this solution, we write
\begin{equation} \label{qpert}
q=q_0(x, t)+\tilde{q}(x, t), 
\end{equation}
where $\tilde{q}\ll 1$. When Eq. (\ref{qpert}) is substituted into the evolution 
equation (\ref{q}) and higher order terms in $\tilde{q}$ neglected, the linearized 
equation is
\begin{equation}
L_{\mbox{\scriptsize kh}}\tilde{q}=0, 
\end{equation}
where $L_{\mbox{\scriptsize kh}}$ is the linearization operator. Here the subscript ``kh'' 
is the abbreviation of the KdV hierarchy. The adjoint operator of $L_{\mbox{\scriptsize kh}}$ will
be denoted as $L_{\mbox{\scriptsize kh}}^A$. We will also denote 
$L_{s0}^+$ and $L_{s0}$ as the operators $L_s^+$ and $L_s$ with
$q(x, t)$ replaced by the solution $q_0(x, t)$.
The objective of this section is to show that
$L_{\mbox{\scriptsize kh}}$ and $L_{s0}^+$ are commutable, and 
$L_{\mbox{\scriptsize kh}}^A$ and $L_{s0}$ are commutable, i.e., 
\begin{equation} \label{kdvcomm1}
L_{\mbox{\scriptsize kh}} L_{s0}^+=L_{s0}^+ L_{\mbox{\scriptsize kh}}, 
\end{equation}
and
\begin{equation} \label{kdvcomm2}
L_{\mbox{\scriptsize kh}}^A L_{s0}=L_{s0} L_{\mbox{\scriptsize kh}}^A. 
\end{equation}
These results are analogous to Eqs. (\ref{comm1}) and (\ref{comm2}) for the AKNS hierarchy. 

Without loss of generality, we will just prove relations (\ref{kdvcomm1}) and (\ref{kdvcomm2})
for power functions of the phase velocity function, $C(z)=z^n$, where $n$ is a non-negative
integer. For this power function, it is easy to check that the linearization operator $L_{\mbox{\scriptsize kh}}$
is: 
\begin{equation}
L_{\mbox{\scriptsize kh}}\tilde{q}=\tilde{q}_t+(4L_{s0}^+)^n\tilde{q}_x
+\sum_{k=1}^n (4L_{s0}^+)^{k-1}{\mathcal{F}}_{\mbox{\scriptsize kh}}[\tilde{q}](4L_{s0}^+)^{n-k}q_{0x}, 
\end{equation}
where 
\begin{equation}
{\mathcal{F}}_{\mbox{\scriptsize kh}}[\tilde{q}]=-4\tilde{q}+2\tilde{q}_x\int_{x}^\infty dy, 
\end{equation}
and $q_0(x, t)$ is a solution of the evolution equation (\ref{q}). Denoting
\begin{equation}
W_n=-(4L_{s0}^+)^n q_{0x}, 
\end{equation}
then $q_{0t}$ is simply 
\begin{equation}
q_{0t}=W_n, 
\end{equation}
where functions $W_n$ satisfy the recursion relation
\begin{equation} \label{Wrecur}
W_{n+1}=4L_{s0}^+ W_n. 
\end{equation}
To show that $L_{\mbox{\scriptsize kh}}$ and $L_{s0}^+$ are commutable, we calculate the quantity
\begin{equation}
{\mathcal{J}}_n \equiv -4(L_{s0}^+L_{\mbox{\scriptsize kh}}-L_{\mbox{\scriptsize kh}}L_{s0}^+)\tilde{q}, 
\end{equation}
which has the expression
\begin{eqnarray}
{\mathcal{J}}_n= & 2W_{nx}\int_x^\infty \tilde{q} dy-4W_n \tilde{q}
-(4L_{s0}^+)^n\left[ 4L_{s0}^+\tilde{q}_x-4(L_{s0}^+\tilde{q})_x\right] \nonumber \\
& -\sum_{k=1}^n (4L_{s0}^+)^{k-1}\left\{4L_{s0}^+ {\mathcal{F}}_{\mbox{\scriptsize kh}}[\tilde{q}]
-{\mathcal{F}}_{\mbox{\scriptsize kh}}[4L_{s0}^+ \tilde{q}]\right\}(4L_{s0}^+)^{n-k}q_{0x}. 
\end{eqnarray}
Now we show that ${\mathcal{J}}_n=0$ for all $n\ge 0$ by the induction method. 

When $n=0$ or 1, trivial calculations show that ${\mathcal{J}}_n$ is indeed zero. 
Now assume that ${\mathcal{J}}_n=0$ for some $n\ge 0$. Notice that 
\begin{eqnarray} \label{Jn+1}
{\mathcal{J}}_{n+1}-4L_{s0}^+ {\mathcal{J}}_n= & 
2W_{n+1, x}\int_x^\infty \tilde{q} dy -4W_{n+1}\tilde{q}
-4L_{s0}^+\left[ 2W_{nx}\int_x^\infty \tilde{q} dy-4W_n\tilde{q}\right]
 \nonumber \\
& +\left\{4L_{s0}^+{\mathcal{F}}_{\mbox{\scriptsize kh}}[\tilde{q}]
-{\mathcal{F}}_{\mbox{\scriptsize kh}}[4L_{s0}^+\tilde{q}]\right\}W_n. 
\end{eqnarray}
Substituting the recursion relation (\ref{Wrecur}) for $W_{n+1}$ into the above equation
(\ref{Jn+1}) and carrying out some algebraic simplifications including integration by parts, 
we find that 
\begin{equation}
{\mathcal{J}}_{n+1}-4L_{s0}^+ {\mathcal{J}}_n= 0. 
\end{equation}
Since ${\mathcal{J}}_n=0$ by assumption, we see that ${\mathcal{J}}_{n+1}=0$. This induction
procedure proves that ${\mathcal{J}}_n=0$ for all $n\ge 0$. Thus, 
$L_{\mbox{\scriptsize kh}}$ and $L_{s0}^+$ are commutable for any power function of the phase velocity
$C(z)$. The commutability for general entire functions of $C(z)$ follows from the fact
that an entire function can be expanded into a power series. 

Now that $L_{\mbox{\scriptsize kh}}$ and $L_{s0}^+$ are commutable. Taking the adjoint of the 
commutability relation (\ref{kdvcomm1}), we find that 
$L_{\mbox{\scriptsize kh}}^A$ and $L_{s0}$ are also commutable. 

Commutability of $L_{\mbox{\scriptsize kh}}$ ($L_{\mbox{\scriptsize kh}}^A$) and $L_{s0}^+$ ($L_{s0}$)
implies that these operators share the same set of eigenfunctions and generalized eigenfunctions. 
The eigenfunctions of $L_{s0}^+$ and $L_{s0}$ are well known \cite{AKNS, newell1980}. 
They are simply squared eigenfunctions of the Schr\"odinger operator with potential $q_0(x, t)$. 
Specifically, consider the Schr\"odinger equation
\begin{equation} \label{schroedinger}
v_{xx}+\left[\zeta^2+q_0(x, t)\right]v=0. 
\end{equation}
Using conventional notation, we define the eigenstates 
$\psi(x, t, \zeta)$ and $\phi(x,t, \zeta)$ of Eq. (\ref{schroedinger}) as
\begin{equation}
\psi(x, t, \zeta) \longrightarrow \left\{ \begin{array}{l}
e^{i\zeta x}, \hspace{1cm}  x \rightarrow \infty; \\
a(t, \zeta)e^{i\zeta x}-b(t, -\zeta)e^{-i\zeta x}, \hspace{0.5cm} x \rightarrow -\infty; 
\end{array}
\right. 
\end{equation}
and
\begin{equation}
\phi(x, t, \zeta) \longrightarrow \left\{ \begin{array}{l}
e^{-i\zeta x}, \hspace{1cm}  x \rightarrow -\infty; \\
a(t, \zeta)e^{-i\zeta x}+b(t, \zeta)e^{i\zeta x}, \hspace{0.5cm} x \rightarrow \infty. 
\end{array}
\right. 
\end{equation}
In addition to the above continuous spectrum (real $\zeta$), Eq. (\ref{schroedinger}) 
may also possess discrete eigenvalues in the upper half $\zeta$-plane (on the
imaginary axis for real potential $q_0$) where $a(t, \zeta_k)=0, k=1, 2, \dots, N$. 
Note that if $q_0(x, t)$ is a solution of the KdV hierarchy (\ref{q}), 
then these discrete eigenvalues $\zeta_k$ are independent of time $t$
\cite{AKNS, GGKM}. With the above notations, 
the set of eigenfunctions and generalized eigenfunctions for the operator $L_{s0}^+$ is
\begin{equation} \label{kdvset1}
\left\{ \left. \frac{\partial \psi^2}{\partial x} \right|_{\zeta}, 
\zeta \; \mbox{real}; 
\left. \frac{\partial \psi^2}{\partial x} \right|_{\zeta_k}, 
\left. \frac{\partial^2 \psi^2}{\partial x\partial \zeta } \right|_{\zeta_k}, 
1\le k \le N \right\}, 
\end{equation}
and the set of such eigenfunctions for $L_{s0}$ is
\begin{equation} \label{kdvset2}
\left\{ \left. \phi^2 \right|_{\zeta}, \zeta \; \mbox{real}; 
\left. \phi^2 \right|_{\zeta_k}, \left. \frac{\partial \phi^2}{\partial \zeta}\right|_{\zeta_k}, 
1\le k \le N \right\}. 
\end{equation}
Commutability of $L_{s0}^+$ ($L_{s0}$) and $L_{\mbox{\scriptsize kh}}$ ($L_{\mbox{\scriptsize kh}}^A$) 
shows that the sets (\ref{kdvset1}) and (\ref{kdvset2}) are also 
eigenfunctions and generalized eigenfunctions for the linearization operators
$L_{\mbox{\scriptsize kh}}$ and $L_{\mbox{\scriptsize kh}}^A$ respectively. 
In addition, we can readily show that the eigenvalue relations are
\begin{equation}
L_{\mbox{\scriptsize kh}}\left. \frac{\partial \psi^2}{\partial x} \right|_{\zeta}=2i\zeta C(4\zeta^2)
\left. \frac{\partial \psi^2}{\partial x} \right|_{\zeta}, \hspace{0.5cm} \zeta\; \mbox{real}; 
\end{equation}
\begin{equation}
L_{\mbox{\scriptsize kh}}\left. \frac{\partial \psi^2}{\partial x} \right|_{\zeta_k}=2i\zeta_k C(4\zeta_k^2)
\left. \frac{\partial \psi^2}{\partial x} \right|_{\zeta_k}, \hspace{0.5cm} 1\le k \le N;  
\end{equation}
\begin{equation}
L_{\mbox{\scriptsize kh}}\left. \frac{\partial^2 \psi^2}{\partial x\partial \zeta } \right|_{\zeta_k}=
2i\zeta_kC(4\zeta_k^2)
\left. \frac{\partial^2 \psi^2}{\partial x\partial \zeta } \right|_{\zeta_k}
+\left[ 2iC(4\zeta_k^2)+16i\zeta_k^2 C'(4\zeta_k^2)\right]
\left. \frac{\partial \psi^2}{\partial x} \right|_{\zeta_k}, 
\hspace{0.5cm} 1\le k \le N;  
\end{equation}
and
\begin{equation}
L_{\mbox{\scriptsize kh}}^A \left. \phi^2 \right|_{\zeta}=2i\zeta C(4\zeta^2)\left. \phi^2 \right|_{\zeta}, 
\hspace{0.5cm} \zeta \; \mbox{real}; 
\end{equation}
\begin{equation}
L_{\mbox{\scriptsize kh}}^A \left. \phi^2 \right|_{\zeta_k}=2i\zeta_k C(4\zeta_k^2)\left. \phi^2 \right|_{\zeta_k}, 
\hspace{0.5cm} 1\le k \le N; 
\end{equation}
\begin{equation}
L_{\mbox{\scriptsize kh}}^A  \left. \frac{\partial \phi^2}{\partial \zeta}\right|_{\zeta_k}
=2i\zeta_k C(4\zeta_k^2)\left. \frac{\partial \phi^2}{\partial \zeta}\right|_{\zeta_k}
+\left[2iC(4\zeta_k^2)+16i\zeta_k^2C'(4\zeta_k^2)\right]\left. \phi^2 \right|_{\zeta_k}, 
\hspace{0.5cm} 1\le k \le N. 
\end{equation}
The completeness of the two sets (\ref{kdvset1}) and (\ref{kdvset2}) and their
inner products have been derived in \cite{newell1980, sachs}. Thus these sets can be used 
to expand the perturbation solutions in a direct soliton/multi-soliton perturbation theory
\cite{herman1990, yan1996, yang2000}. 

\section{Concluding remarks}
In this article, we have studied the linearization operators of the AKNS hierarchy and 
KdV hierarchy equations expanded around an arbitrary solution. We have found that
these linearization operators and the recursion operator which 
generates the hierarchy are commutable. This commutability relation immediately
reveals that linearization operators and the recursion operator
share the same set of eigenfunctions, and these eigenfunctions are simply 
squared eigenfunctions of the Zakharov-Shabat or Schr\"odinger equations. 
Compared to the other methods for determining eigenfunctions of the linearization
operators \cite{keener1977, herman1990}, our method is simple, and
it gives the eigenfunctions for the entire AKNS and KdV hierarchies all at once. 
In addition, our result makes a clear connection between the direct
soliton/multi-soliton perturbation theory and the inverse-scattering based perturbation theory, 
as perturbation solutions in both theories are expanded onto the same complete set of 
eigenfunctions. 
With the eigenfunctions of linearization operators now available, one can proceed to 
develop a direct soliton/multi-soliton perturbation theory for the AKNS and KdV 
hierarchies, which should reproduce the results of \cite{kaupnewell1978, newell1980}
obtained by the inverse-scattering based perturbation method. This problem falls outside
the scope of the present article. Another interesting question is whether the idea 
of this paper can be extended to derive eigenfunctions of linearization operators
for other integrable equations. We believe the answer is positive. 
For instance, our results should be extendible to the integrable
$N$-wave equations whose recursion operator and its eigenfunctions 
have been available \cite{zakharovbook, gerdjikov}. 
This question will be left for future studies. 

\section*{\hspace{0.1cm} Acknowledgments}
The author thanks Prof. D.J. Kaup for very helpful discussions on this subject. 
This work was supported in part by 
the Air Force Office of Scientific Research under contract number USAF F49620-99-1-0174, 
and by National Science Foundation under grant number DMS-9971712.

\end{document}